\documentclass[12pt]{article}
\font\smallit=cmti10

\begin{document}

\begin{center}
{\bf ANOTHER PROOF OF AN EXTENSION OF A CURIOUS IDENTITY} \vskip
20pt
{\bf Yidong Sun}\\
{\smallit Department of Applied Mathematics,
Dalian University of Technology, Dalian 116024, P. R. China}\\
{\tt sydcom@eyou.com}\\  \vskip 10pt
\end{center}

\centerline{\bf Abstract}

\noindent Based on Jensen formulae and the second kind of
Chebyshev polynomials, another proof is presented for an extension
of a curious binomial identity due to Z. W. Sun and K. J. Wu.

\thispagestyle{empty} \baselineskip=15pt \vskip 30pt

\section*{\normalsize 1. Introduction}
Recently, Z. W. Sun and K. J. Wu \cite{SW} have presented an
extension of a curious binomial identity \cite{ZWS}, that is the
following generalized version:
\begin{eqnarray}\label{eqn 1.1}
\lefteqn{(x+(m+1)z)\sum_{k=0}^m(-1)^{k}{{x+y+kz}\choose{m-k}}{{y+k+kz}\choose{k}}}\nonumber\\[5pt]
&=&z\sum_{0\leq i\leq k\leq m}(-1)^k{k\choose
i}{{x+i}\choose{m-k}}(1+z)^{k+i}(1-z)^{k-i}+(x-m){x\choose m}.
\end{eqnarray}
In 2002 Z. W. Sun \cite{ZWS} proved the case when $z=1$ using
double recursion. Later four alternative proofs have been provided
for the special case. A generating function proof was given by A.
Panholzer and H. Prodinger \cite{PP}; D. Merlini and R. Sprugnoli
\cite{MS} established it through Riordan arrays; S. B. Ekhad and
M. Mohammed \cite{EM} proved it based on a WZ method. Later, W.
Chu and L. V. D. Claudio \cite{CC} reobtained the identity by
using Jensen formulae \cite{C0}. More recently, D. Callan
\cite{Callan} gave a combinatorial proof using weight-reversing
involutions on suitable configurations involving dominos and
colorings.

Motivated by the method \cite{CC}, we give a simple proof of the
identity (\ref{eqn 1.1}). \vskip 30pt

\section*{\normalsize 2. Proof of (\ref{eqn 1.1})}
Let
\begin{eqnarray*}
f(x,y,z)&=&\sum_{k=0}^m(-1)^{k}{{x+y+kz}\choose{m-k}}{{y+k+kz}\choose{k}}
\hskip0.5cm {\rm and}\\
g(x,z)&=&\sum_{0\leq i\leq k\leq m}(-1)^k{k\choose
i}{{x+i}\choose{m-k}}(1+z)^{k+i}(1-z)^{k-i}.
\end{eqnarray*}
Recall the Jensen formulae on binomial convolutions
\begin{eqnarray*}
\sum_{i=0}^m{{a+bi}\choose
i}{{c-bi}\choose{m-i}}=\sum_{j=0}^m{{a+c-j}\choose {m-j}}b^j,
\end{eqnarray*}
and the second kind of Chebyshev polynomials
\begin{eqnarray*}
U_n(t)&=&\frac{sin(n+1)\theta}{sin(\theta)}=\sum_{k=0}^{[n/2]}(-1)^{k}{{n-k}\choose{k}}(2t)^{n-2k},
\hskip0.3cm {\rm by \ setting} \ t=cos(\theta).
\end{eqnarray*}
Then we have
\begin{eqnarray*}
f(x,y,z)&=&\sum_{k=0}^m(-1)^{k}{{x+y+kz}\choose{m-k}}{{y+k+kz}\choose{k}}\\
    &=&(-1)^m\sum_{k=0}^m{{y+k(z+1)}\choose{k}}{{-1-x-y+m-k(z+1)}\choose{m-k}}\\
    &=&(-1)^m\sum_{j=0}^m{{-1-x+m-j}\choose{m-j}}(1+z)^j\\
    &=&\sum_{j=0}^m{{x}\choose{m-j}}(-1-z)^j.
\end{eqnarray*}
Now we can express $g(x,z)$ as a simple form
\begin{eqnarray*}
g(x,z)&=&\sum_{k=0}^m\sum_{i=0}^k(-1)^k{k\choose
i}{{x+i}\choose{m-k}}(1+z)^{k+i}(1-z)^{k-i}\\
&=&\sum_{k=0}^m\sum_{i=0}^k(-1)^k(1+z)^{k+i}(1-z)^{k-i}
\sum_{j=k}^m{{x}\choose{m-j}}{{i}\choose{j-k}}{k\choose i}\\
&=&\sum_{j=0}^m{{x}\choose{m-j}}\sum_{k=0}^j\sum_{i=0}^k(-1)^k(1+z)^{k+i}(1-z)^{k-i}
{{i}\choose{j-k}}{k\choose i}\\
&=&\sum_{j=0}^m{{x}\choose{m-j}}\sum_{k=0}^j\sum_{i=0}^k(-1)^k(1+z)^{k+i}(1-z)^{k-i}
{{k}\choose{j-k}}{{2k-j}\choose {k+i-j}}\\
&=&\sum_{j=0}^m{{x}\choose{m-j}}(1+z)^j\sum_{k=0}^j(-1)^k{{k}\choose{j-k}}\sum_{i=0}^k{{2k-j}\choose
{k+i-j}}(1+z)^{k+i-j}(1-z)^{k-i}\\
&=&\sum_{j=0}^m{{x}\choose{m-j}}(1+z)^j\sum_{k=0}^j(-1)^k{{k}\choose{j-k}}2^{2k-j}\\
&=&\sum_{j=0}^m{{x}\choose{m-j}}(-1-z)^jU_j(t)|_{t\rightarrow 1}\\
&=&\sum_{j=0}^m(j+1){{x}\choose{m-j}}(-1-z)^j.
\end{eqnarray*}
Hence we obtain
\begin{eqnarray*}
\lefteqn{(x+(m+1)z)f(x,y,z)-zg(x,z)}\\
&=&\sum_{j=0}^m(x+(m+1)z-(j+1)z){{x}\choose{m-j}}(-1-z)^j\\
&=&\sum_{j=0}^m(x-m+j){{x}\choose{m-j}}(-1-z)^j-(m-j){{x}\choose{m-j}}(-1-z)^{j+1}\\
&=&\sum_{j=0}^m(1+m-j){{x}\choose{1+m-j}}(-1-z)^j-(m-j){{x}\choose{m-j}}(-1-z)^{j+1}\\
&=&(1+m){{x}\choose{1+m}}=(x-m){{x}\choose{m}},
\end{eqnarray*}
as desired (the second to last sum is telescoping). \vskip 30pt

\renewcommand{\refname}{{\normalsize References}}

\end{document}